\documentclass[a4paper]{article}
\usepackage{latexsym}
\begin{document}

\newtheorem{definition}{Definition}
\newtheorem{conjecture}{Conjecture}
\newtheorem{theorem}{Theorem}
\newtheorem{proposition}{Proposition}
\newtheorem{remark}{Remark}
\newtheorem{corollary}{Corollary}
\newtheorem{lemma}{Lemma}
\newtheorem{observation}{Observation}

\newtheorem{hconjecture}{Conjecture}

\newcommand{\qed}{\hfill$\Box$\medskip}

\title{Busy Beaver Scores and Alphabet Size} 
\author{Holger Petersen\\  
Reinsburgstr. 75\\
70197 Stuttgart\\
Germany} 

\maketitle

\begin{abstract}
We investigate the Busy Beaver Game
introduced by Rado (1962) generalized to non-binary alphabets. 
Harland (2016) conjectured that activity (number of steps) 
and productivity (number of non-blank symbols) of candidate 
machines grow as the alphabet size increases. 
We prove this conjecture for any alphabet size under the 
condition that the number of states is sufficiently large.
For the measure activity we show that increasing the 
alphabet size from two to three allows an increase.
By a classical construction it is even possible to obtain 
a two-state machine increasing activity and productivity 
of any machine if we allow an alphabet size depending on 
the number of states of the original machine.  
We also show that an increase of the alphabet by 
a factor of three admits an increase of activity.
\end{abstract}

\section{Introduction}
The Busy Beaver Game, as originally defined by Rado \cite{Rad62},
is to determine for a given number $n$ of states of deterministic Turing machines
over the alphabet $\{0,1\}$ ($0$ is the blank symbol) 
the maximum number of ones produced on an initially
blank two-way inifinite tape.
In each step such a machine reads a tape symbol and depending on
the current state writes a symbol, shifts
its head one square to the left or to the right, and enters a new
state. There is a single halt state (which is traditionally not
counted), and on the transition to this
state the machine also writes a symbol. 
What we have just described is sometimes called the {\em quintuple variant} 
of Turing machines in view of the five pieces of information
that define a transition. In contrast, the {\em quadruple variant}
can either move the tape head or write a symbol but not both.

Rado introduced the function $\Sigma(n)$ as the maximum
(number of ones produced by machines with $n$ states. The function
$S(n)$ denotes the maximum number of steps performed (shift-number) 
of such machines. He proved that these functions 
are non-computable and even grow faster than any computable function.
Rado also pointed out that these are very simple examples of non-computable
functions and that no (explicit) enumeration of computable
functions is used in their definition.

The functions are of metamathematical interest as well, since 
open problems like Goldbach's conjecture, which can be refuted 
in a constructive way by a counterexample, would be settled if
$S(n)$ would be computable for an $n$ large enough to
determine a counterexample by running a Turing machine
 \cite{Bra88,Cha87}. Recently explicit bounds on such an 
$n$ have been determined for Goldbach's conjecture 
and the Riemann hypothesis along with a Turing machine 
that cannot be proved to run forever in ZFC \cite{YA16}.

Here we consider the generalization of the Busy Beaver Game
to alphabets with more than two symbols. As in \cite{Mic16} we 
denote by $\Sigma(n, m)$ the maximum number of non-blanks
produced by any halting deterministic Turing machine with $n$ states and
$m$ symbols (called {\em productivity}) workin on an initially blank tape.
Similarly, we denote by $S(n, m)$ the maximum number of steps performed
(called {\em activity}). 
Thus the functions defined by Rado are now special cases with $m = 2$.
For a specific Turing machine $M$ we denote the
two measures by $\mbox{productivity}(M)$ and $\mbox{activity}(M)$. 

A Turing machine $M$ participating in the generalized
Busy Beaver competition can be represented by a table of the form
\begin{center}
\begin{tabular}{cc|c|c|c|c|c}
&\multicolumn{1}{c}{}  & \multicolumn{4}{c}{input symbol}\\[1ex]
&\multicolumn{1}{c}{} &  \multicolumn{1}{c}{0} & \multicolumn{1}{c}{1} & \multicolumn{1}{c}{$\cdots$}  & \multicolumn{1}{c}{$m-1$} \\[1ex]\cline{3-6}
& & &  & &  \\[-1ex]
current state
&1   &   $w^0_1\delta^0_1s^0_1$       & $w^1_1\delta^1_1s^1_1$   &  $\cdots$   & $w^{m-1}_1\delta^{m-1}_1s^{m-1}_1$  &  \ \ \ \ \ \ \ \ \\[1ex]
&2   &   $w^0_2\delta^0_2s^0_2$       & $w^1_2\delta^1_2s^1_2$   &  $\cdots$   & $w^{m-1}_2\delta^{m-1}_2s^{m-1}_2$ & \\[1ex]
&\vdots   &   \vdots       &   \vdots       & $\cdots$   &  \vdots  & \\[1ex]
&$n$   &   $w^0_n\delta^0_ns^0_n$       & $w^1_n\delta^1_ns^1_n$   &  $\cdots$   & $w^{m-1}_n\delta^{m-1}_ns^{m-1}_n$ & \\[1ex]\cline{3-6}
\end{tabular}
\end{center}
where $w^i_k\in\{0, 1, \ldots, {m-1} \}$ indicates the symbol written by $M$ after
reading $i$ in state $k$, $\delta^i_k\in\{L, R\}$ is the direction of
the head movement, and $s^i_k\in\{1, \ldots, n+1\}$ is the new state $M$
enters. State~$1$ is the initial state and state~$n+1$ is the halting state.

As early as 1966, the lower bounds 
$\Sigma(3, 3) \ge 12$ and $S(3, 3) \ge 57$
were reported in \cite{Kor66} for a non-binary alphabet%
\footnote{The origin of these bounds
communicated to Korfhage by C.~Y.~Lee of Bell Telephone Lab.\ is not clear. In
\cite{Kor66} Lee, Tibor Rado, Shen Lin, Patrick Fischer, Milton Green, 
and David Jefferson are mentioned in connection with these lower bounds and other
early results.}.
Over the following decades, investigations concentrated on 
computing $\Sigma(4, 2)$ and on improving lower bounds for larger numbers 
of states in the classical setting of a binary tape alphabet.
The progress in the chase of Busy Beavers is
reflected in the following table:
\begin{center}
\begin{tabular}{|c|c|c|c|}\hline
$n$ & $\Sigma(n, 2)$ &  $S(n, 2)$ & references \\\hline
1   &    1    &   1       &  Rado \cite{Rad62} \\\hline
2   &    4    &   6       &  Rado \cite{Rad62} \\\hline
3   &    6    &   21      &  Lin, Rado \cite{LR65} \\\hline
4   &    13   &   107    &   Brady \cite{Bra83} \\\hline
5   &  $\ge$4098   &  $\ge$47,176,870 & Marxen, Buntrock \cite{MB90}\\\hline
6   & $\ge 3.514 \cdot 10^{18,267}$ & $\ge 7.412 \cdot 10^{36,534}$ & Kropitz, see \cite{MBweb}\\\hline
\end{tabular}
\end{center}

With the exception of lower bounds due to Brady 
($S(2,3) \ge 38$, $S(2,4) \ge 7,195$),
the search for high scoring machines with more than two symbols did not 
continue before 2004. As outlined in the survey \cite{Mic16}, 
Michel and Brady improved the lower bounds on $\Sigma(3,3)$ and $S(3,3)$
during that year. Between 2005 and 2008 many new machines for non-binary 
alphabets were found mainly by two teams: Gr\'{e}gory Lafitte and
Christophe Papazian, and Terry and Shawn Ligocki (father and son).
Lafitte and Papazian could also establish that $\Sigma(2,3) = 9$ and
$S(2,3) = 38$, confirming Michel's conjecture from \cite{Mic04}
that Brady's lower bounds dating back almost two decades were tight. 

Given the known values and lower bounds for non-binary 
alphabets, it is natural expect that $\Sigma(n, m)$ and 
$S(n, m)$ are increasing in both parameters (it is easily shown
that they are increasing in their first parameter, see Lemma~\ref{add_state} 
below). An even stronger conjecture was stated by Harland \cite{Har16}.

Before presenting our results we cite Harland's conjecture:
\setcounter{hconjecture}{27}
\begin{hconjecture}[in \cite{Har16}]\label{harland}
Let $M$ be a $k$-halting Turing machine
with $n$ states and $m$ symbols for some $k \ge 1$ with
finite activity. Then there is a $k$-halting $n$-state $(m + 1)$-symbol 
Turing machine $M'$ with 
finite activity such that $$\mbox{activity}(M') > \mbox{activity}(M)
\mbox{ and } \mbox{productivity}(M') > \mbox{productivity}(M).$$
\end{hconjecture}
Here $k$-halting means that there are $k$ transitions to the halting state.

For $n = 1$, an $n$-state Turing machine has to halt after the 
first step on a blank in order to have finite activity. As this holds
independently of the size of the alphabet, no increase
of activity and productivity is possible. We therefore exclude the
trivial case $n = 1$.

Notice that the conjecture is stronger than just stating that
$\Sigma$ and $S$ are increasing as $m$ grows and $n$ is kept fixed
(which it implies by taking highest scoring machines as $M$).
The conjecture considers for any {\em specific} machine
{\em both} activity and productivity at the same time. 
A machine maximizing one of the measures may in fact
not maximize the other, as is the case for $n =3$ where
machines with activity 21 produce at most $5 < \Sigma(3)$
ones. 
 
In addition Harland's conjecture
imposes a restriction on the structure of 
a machine increasing these measures, namely that the number of
halting transitions is kept constant for machine $M'$.

Highest scores
for small machines still provide evidence in support 
of the conjecture. We have 
$$\Sigma(2, 2) = 4 < \Sigma(2, 3) = 9 < 2,050 \le \Sigma(2, 4),$$
$$S(2, 2) = 6 < S(2, 3) = 38 < 3,932,964 \le S(2, 4),$$
$$\Sigma(3, 2) = 6 < 374,676,383 \le \Sigma(3, 3),$$
and 
$$S(3, 2) = 21 < 119,112,334,170,342,540 \le S(3, 3)$$
(results of Rado, Lin, Lafitte, Papazian, T.~Ligocki and S.~Ligocki,
see \cite{Mic16} for references).

\section{Results}

It is well-known that activity and 
productivity grow with the number of states, see the 
figure on p.~77 of \cite{Hop84} or Proposition~27 of
\cite{Har16}.
\begin{lemma}\label{add_state}
Let $M$ be a Turing machine
with $n$ states and $m$ symbols with
finite activity. Then there is an $(n+1)$-state $m$-symbol 
Turing machine $M'$ with 
finite activity
such that $\mbox{activity}(M') > \mbox{activity}(M)$ 
and $\mbox{productivity}(M') > \mbox{productivity}(M)$.
\end{lemma}
The lemma can be proved for any alphabet by 
redirecting the (unique) halting transition to the 
new state and having it
skip symbols different from the blank while moving
the head in one direction. The first blank encountered 
is replaced with a non-blank and then the machine halts.

An encoding scheme originally developed by Ben-Amram and Petersen \cite{BP02}
and called {\em introspective computing} 
by Luke Schaeffer \cite{YA16} will be essential
in proving Harland's conjecture for sufficiently large numbers of states.
\begin{theorem}\label{limit}
For every $m \ge 2$ and $k \ge 1$ there is an 
$N_{m,k}$ such that for every $k$-halting Turing machine 
$M$ with $n \ge N_{m,k}$ states and $m$ symbols with 
finite activity there is an $n$-state, 
$(m+1)$-symbol $k$-halting Turing machine $M'$ with finite activity
such that $\mbox{activity}(M') > \mbox{activity}(M)$ 
and $\mbox{productivity}(M') > \mbox{productivity}(M)$.
\end{theorem} 
Proof. Let $M$ be a Turing machine as described in the 
theorem with $n \ge m$ states. 
We first notice that w.l.o.g. all $n$ states
appear in the unique halting computation of $M$
on the blank tape. For otherwise we omit an
unused state $s$ (reducing the number of halting transitions
by at most $m$) and redirect all transitions with target $s$
to some remaining state. The resulting Turing machine 
$\hat M$ with $n-1$ states is equivalent to $M$ on a blank
tape, since none of the modified transitions is ever reached
in the course of the computation. 
We apply Lemma~\ref{add_state}
to $\hat M$ resulting in a machine $M'$ with 
$\mbox{activity}(M') > \mbox{activity}(M)$ and 
$\mbox{productivity}(M') > \mbox{productivity}(M)$.
Since the construction for Lemma~\ref{add_state} 
preserves the number of halting transitions, it suffices to add 
at most one halting transition on the new symbol $m$
for each state in order to transform $M'$ into a 
$k$-halting machine. These transitions will not influence
the computation because symbol $m$ is never written
onto the tape. In the following we let $N_{m,k} \ge m$.

The next normalization of $M$ is the observation from \cite{BP02}
that in its computation on a blank tape ``new'' states (states not
previously visited) appear in increasing order, i.e., the first
state visited and not in the set $\{ 1,\ldots, s\}$ is $s+1$.
This can be achieved by renaming the states appearing in
the unique computation of $M$ on a blank tape.
A transition followed when a state $s$ is first arrived at
is called {\em special}, all other transitions are {\em ordinary}.
Targets of special transitions can be omitted
from a description of $M$,
as long as there is a flag indicating whether a transition is special.
We further note that the number of special transitions is exactly $n$, since
by the normalization above all states (including the halt state)
are reached.

Finally halting tansitions (except the one appearing in the
halting computation) are modified, such that they target another state.
Obviously this does not influence the computation.

After these transformations, $M$ can be described by the following information:
\begin{enumerate}
\item
The number $n-1$ in a self-delimiting binary notation, using at most
$2\lceil\log_2 n\rceil$ bits.
\item
An array containing $m(\lceil\log_2 m\rceil + 2)$ 
bits for every state $i\in \{1,\ldots, n\}$.
These bits correspond to the components (symbol written, 
head movement, and new state) of a row of the transition table
encoding all transitions from a state. The next state is replaced
by a flag that is 1 if and only if the transition is special.
\item
A list of $n(m-1)$ destinations of ordinary transitions. The list is
sorted according to their first appearance in the computation
on a blank tape. A destination can be encoded in 
$\lceil\log_2 n\rceil$ bits, since the halting transition is always
special and the halting state does not appear in another transition.
\end{enumerate}
In summary, the description of $M$ requires
$nm\lceil\log_2 n\rceil - n\lceil\log_2 n\rceil + cn$ bits 
for some constant $c$ if $m$ is fixed.

Next we consider the information content of $n'$ states 
acting as a ROM in the finite control of 
a Turing machine with $m+1$ symbols. By the technique
of introspective computing \cite{BP02} generalized to $m+1$ tape 
symbols, $n'm\lfloor\log_2 n'\rfloor$
bits can be extracted from these states by a fixed extractor
machine $E$ with $n_E$ states. The extracted bits can be processed
by a universal Turing machine $U$ having $n_U$ states and
simulating machines with $m$ symbols. As opposed to usual
simulators, we let $U$ write an extra non-blank symbol after it
has reached the halting transition of the machine being simulated 
(notice that this will make sure that activity as well as productivity
increase in comparison to $M$).
A further specific requirement is that $U$ keeps track
of the first appearance of a state and finalizes the transition table
according to the flags while simulating a machine.
Finally an ordinary universal Turing machine would have exactly
one halting transition. In order to satisfy the requirements
of Harland's conjecture we add a sufficient number of 
(unreachable) states to accommodate $k-1$ additional halting 
transitions.

We let $d = n_E + n_U$, $n' = n - d$ and observe that 
$n'm\lfloor\log_2 n'\rfloor = (n-d)m\lfloor\log_2 n-d\rfloor 
\ge (n-d)m(\lfloor\log_2n\rfloor-1) 
\ge nm\lfloor\log_2n\rfloor- dm\lfloor\log_2n\rfloor - nm + dm 
\ge nm\lceil\log_2n\rceil - dm\lfloor\log_2n\rfloor - 2nm + dm
\ge nm\lceil\log_2 n\rceil - n\lceil\log_2 n\rceil + cn$
for $n \ge N_{m,k}$ with a sufficiently large $N_{m,k}$. 
Therefore $n'$ states suffice to encode $M$.

Finally we compose the
Turing machine over $m+1$ symbols 
with $n'$ states encoding machine $M$,
the extractor $E$, and the universal Turing machine $U$
to obtain machine $M'$ with $n$ states simulating $M$ and 
satisfying the theorem. \qed

Next we consider weaker versions of Harland's conjecture.
But first we show some technical Lemmas.

\begin{lemma}\label{exceedn}
For all $n, m \ge 2$ we have $S(n, m) > n$
\end{lemma} 
Proof. $S(2, 2) = 6 > 2$, Suppose 
$S(n, 2) > n$ for some $n \ge 2$. 
By Lemma~\ref{add_state} we get $S(n+1, 2) > n+1$ and 
$S(n, m) \ge S(n, 2) > n$ by adding transitions 
on $m - 2$ symbols for a two-symbol champion.\qed

\begin{lemma}\label{onedir}
If all transitions of Turing machine $M$
with $n$ states on the blank move the head in the same
direction and $M$ has finite activity, then we have
$\mbox{activity}(M) \le n$.
\end{lemma} 
Proof. If $M$ makes more than $n$ steps in one direction, 
then a state repeats and $M$ does not stop. \qed

The next result is inspired by the construction
in Figure~14 of \cite{Har16}. In contrast to
Theorem~\ref{limit} it does not preserve the
number of halting transitions.
\begin{theorem}\label{alphtwo}
For every Turing machine $M$ with $n\ge 2$ states and two 
symbols having finite activity there is an 
$n$-state, three-symbol Turing machine $M'$ with finite 
activity such that $\mbox{activity}(M') > \mbox{activity}(M)$.
\end{theorem} 
Proof. Without loss of generality $M$ has maximum
activity among all $n$ state, two 
symbol Turing machines and the first transition
of $M$ moves the head to the right. 

We let $M'$ have the basic structure of
$M$ and add transitions on the new (third)
symbol to every state. For a state $s$ to be 
determined below this transition is halting,
while the other transitions are non-halting and
can otherwise be arbitrary, since they 
never will be used.

Consider the tape cell $i$ at the final 
position of the head in the
computation of $M$ on a blank tape. 
We modify the halting transition taken by $M$
to write the new symbol and move the head 
depending on the symbols in neighboring 
cells of $i$.

If cell $i-1$ contains a blank, we modify the halting
transition to move left and go to 
the initial state. By the normalization of
the first transition, $M'$ will move right on the blank
(it cannot halt due to Lemma~\ref{exceedn})
to a state which is chosen as $s$. Then $M$
halts on the new symbol increasing activity by two.

If $i-1$ contains
$1$ and there is a state with a transition moving right
on $1$, we modify the last transition to move left and 
go to such a state. This will increase activity by one
if the transition moving right on $1$ is halting, in which 
case we chosen the current state as $s$.
Otherwise activity increases by two as in the previous case
if $M'$ returns to cell $i$ in a state chosen as $s$.

If all transitions move left
on $1$, we consider tape cell $i+1$. If it contains $1$, we 
modify the halting transition to move right and go to an
arbitrary state. Machine $M'$ will either halt immediately or
return to cell $i$ in a state chosen as $s$ and halt.

Finally consider a blank in cell $i+1$. Since 
for $n \ge 2$ there is a machine with activity exceeding $n$ by
Lemma~\ref{exceedn},
we conclude from Lemma~\ref{onedir} that at least
one transition moves the head left on a blank. Go to a state
with such a transition and move the head to the right.
The resulting Turing machine will halt either when reading
cell $i+1$ or when it returns to cell $i$ in a state chosen as $s$. 

In each case 
$\mbox{activity}(M') \in \{ \mbox{activity}(M)+1, \mbox{activity}(M)+2 \}$.
\qed

Next we turn to constructions that increase the alphabet by more than
one symbol.
\begin{theorem}
For every Turing machine $M$ with $n\ge 2$ states and $m \ge 2$ 
symbols having finite activity there is a 2-state, $(4nm+5m)$-symbol 
Turing machine $M'$ with finite activity
such that $\mbox{activity}(M') > \mbox{activity}(M)$
and $\mbox{productivity}(M') > \mbox{productivity}(M)$.
\end{theorem} 
Proof. Let $M$ be a Turing machine with $n$ states and $m$
symbols. By Lemma~\ref{add_state} there is a machine $M'$
with $n+1$ states and $m$ symbols increasing activity and productivity.
The classical construction from \cite{Sha56} transforms it 
into an equivalent 2-state machine with $4m(n+1)+m = 4nm+5m$ 
symbols. \qed

\begin{theorem}
For every Turing machine $M$ with $n\ge 2$ states and $m \ge 3$ 
symbols having
finite activity there is an $n$-state, $3m$-symbol 
Turing machine $M'$ with finite activity
such that $\mbox{activity}(M') > \mbox{activity}(M)$.
\end{theorem}
Proof. If among the Turing machines with $n$ states and $m$ 
symbols $M$ does not have maximum activity, we choose as 
$M'$ such a machine and no increase of the tape alphabet is
necessary. 

Otherwise for every symbol $a$ of $M$ we add new symbols $a_L$ and
$a_R$ to the transition table of $M'$. 
A transition of $M$ on an old symbol writing $a$
is modified to write $a_R$ if it moves the head to the
left (indicating that $a_R$ is to the right of the
tape head) and similarly $a_L$ if it moves the head to the right. 
On new symbols $a_R$ and $a_L$ machine $M'$ replaces the new
symbol with $a$ and ``bounces'' back to the right if
the symbol was $a_L$ and to the left on $a_R$. Observe that
all symbols with subscript $L$ are to the left of the tape head
or under it and all symbols with subscript $R$ are to the right 
of the tape head or under it in the course of the computation of
$M'$.

Consider the homomorphism $h$ defined by $h(a) = h(a_L) = h(a_R) = a$
for all symbols of $M$. We claim that for every instantaneous
description of $M$ at step $k$ with a tape inscription $w$ 
of cells visited by $M$ and its head on cell $i$ there is an
instantaneous description of $M'$ at step $k' \ge k$ 
with a tape inscription $w'$ satisfying $h(w) = h(w')$ with its 
head on cell $i$. This clearly holds for step 0 when there are
no modified cells. If $M'$ reads
an old symbol $a$ it writes some $b_R$ or $b_L$ while $M$ writes
$b$ and both move their heads in the same direction. This clearly 
maintains the property $h(w) = h(w')$ and that the head positions
correspond. If $M'$ reads $a_L$ it has just moved its head left 
and the neighboring cell contains some symbol $b_R$. Now $M'$ 
writes $a$, moves right, replaces $b_R$ with $b$, and returns
to $a$. In comparison to $M$ two additional steps have been 
performed while $h(a_Lb_R) = h(ab)$. In the same way $M'$ 
behaves on $a_R$. We conclude that $M'$ halts if and only $M$ 
does and $\mbox{activity}(M') \ge \mbox{activity}(M)$.

To see that $\mbox{activity}(M') > \mbox{activity}(M)$ we make use of
the assumption that $M$ has maximum activity among the 
Turing machines with $n$ states and $m$ 
symbols. By Lemma~\ref{exceedn} and Lemma~\ref{onedir} $M'$ has to make 
at least one turn, which adds at least two steps to the computation
of $M'$ in comparison to $M$. \qed

\section{Discussion}
We have partially proved Harland's conjecture. It holds for 
$n$ sufficiently large and (restricted to the measure activity and
without maintaining the number of halting transitions) for $m=2$.
An increase of the alphabet size exceeding one admits similar 
results for all $n$. In the former construction we have used 
the technique of interpretation instead of instrumentation
(in terms of \cite{BP02}). 
 
If the Harland's conjecture is true in general, then it provides further 
evidence for the symmetry of symbols and states discussed by Shannon
in the concluding remarks of \cite{Sha56}, since an increase in one of 
the parameters adds power to the machines.


\begin{thebibliography}{10}
\bibitem{BP02}
Amir M. Ben-Amram and Holger Petersen.
\newblock Improved Bounds for Functions Related to Busy Beavers.
\newblock {\em Theory of Computing Systems}, 35 (1), 2002, 1–-11.

\bibitem{Bra83}
Allen H. Brady.
\newblock The Determination of the Value of {R}ado's Noncomputable Function $\Sigma(k)$
for Four-State {T}uring Machines.
\newblock {\em Mathematics of Computation}, 40 (162), 1983, 647--665.

\bibitem{Bra88}
Allen H. Brady.
\newblock The Busy Beaver Game and the Meaning of Life.
\newblock in: {\em The Universal Turing Machine: A Half-Century Survey}, 2nd Edition, R.~Herken (Ed.), Springer, 
1995, 237–-254.

\bibitem{Cha87}
Gregory Chaitin. 
\newblock Computing the Busy Beaver Function.
\newblock In {\em Open Problems in Communication and Computation}, Springer, 1987, 108--112.

\bibitem{Har16}
James Harland. 
\newblock Generating Candidate Busy Beaver Machines (Or How to Build the Zany Zoo).
\newblock {\tt https://arxiv.org/abs/1610.03184v1}, 2016. 

\bibitem{Hop84}
John E. Hopcroft.
\newblock Turing Machines.
\newblock {\em Scientific American}, May 1984, 250 (5), 70--80.

\bibitem{Kor66}
Robert R.~Korfhage.
\newblock Logic and Algorithms: With Applications to the Computer and Information Sciences, 
New York, Wiley, 1966.

\bibitem{LR65}
Shen Lin and Tibor Rado.
\newblock Computer Studies of {T}uring Machine Problems.
\newblock {\em Journal of the Association for Computing Machinery},
  12, 1965, 196--212.

\bibitem{MB90}
Heiner Marxen and J{\" u}rgen Buntrock.
\newblock Attacking the {B}usy {B}eaver 5.
\newblock {\em Bulletin of the European Association for Theoretical Computer
  Science (EATCS)}, 40, 1990, 247--251.

\bibitem{MBweb}
Heiner Marxen.
\newblock Currently Known Results. (Download Apr 25, 2017)
\newblock {\tt http://www.drb.insel.de/}$\sim${\tt heiner/BB}.

\bibitem{Mic04}
Pascal Michel.
\newblock Small {T}uring Machines and Generalized Busy Beaver Competition.
\newblock {\em Theoretical Computer Science}, 326, 2004, 45--56.

\bibitem{Mic16}
Pascal Michel.
\newblock The Busy Beaver Competition: A Historical Survey.
\newblock {\tt https://arxiv.org/abs/0906.3749v4}, 2016.

\bibitem{Rad62}
Tibor Rado.
\newblock On Non-computable Functions.
\newblock {\em The Bell System Technical Journal}, 41, 1962, 877--884.

\bibitem{Sha56}
Claude E. Shannon.
\newblock A Universal Turing Machine with Two Internal States. 
\newblock In {\em Automata Studies (AM-34)}, edited by C. E. Shannon and J. McCarthy, 
Princeton University Press, 1956, 157-–166.

\bibitem{YA16}
Adam Yedidia and Scott Aaronson.
\newblock A Relatively Small Turing Machine Whose Behavior Is Independent of Set Theory. 
\newblock {\em Complex Systems}, 25 (4), 2016.
{\tt http://www.complex-systems.com/issues/25-4.html}.
\end{thebibliography}
\end{document}